\documentstyle[eqsecnum,aps]{revtex}

\tighten
\begin{document}
\draft
\twocolumn[\hsize\textwidth\columnwidth\hsize\csname
@twocolumnfalse\endcsname

\title{Resistivity of Doped Two-Leg Spin Ladders}

\author{Gilson Carneiro $^{\diamond}$ and Pascal
Lederer$^+$\footnote{On leave from 
Physique des Solides,U. P. S., F91405 Orsay, France( Laboratoire
associ\'e au CNRS)}  
 } 
\address{ $^+$Departamento de F\'{\i}sica, PUC-Rio, C. P. 38071, Rio de
Janeiro and $^{+, \diamond}$Instituto de F\'{\i}sica, Universidade Federal
do Rio de Janeiro, C.P.68528, 21945-970, Rio de Janeiro, Brasil} 
\date{\today}
\maketitle
\begin{abstract}
In doped two-leg spin ladder systems, holes are expected to form charged
bosonic pairs. We study charge transport in this system in the
temperature range where the bosons can be described as weakly
interacting quasi-particles. We consider boson-phonon and boson-impurity
scattering processes. We suggest that due to the Ioffe-Regel
resistivity saturation mechanism, the resistivity may exhibit a local
maximum at intermediate temperatures. We propose that this may explain
a similar feature found in recent experimental results.  
\end{abstract}
\vskip2pc]

\section{ INTRODUCTION}
\label{sec.int}

One of the most exciting advances in the field of high $T_c$
superconductivity  and in the theory of strongly correlated electrons
in the last few years has been  
the experimental study and the theoretical understanding of pure and
doped quantum spin ladders\cite{dagrice}. In contrast to the one
dimensional (1D) antiferromagnetic (AF)
Heisenberg chain, where the ground-state has an infinite correlation
length, and exhibits a slow decay of the spin correlations, even -leg
(undoped) ladders have spin-liquid ground-states with purely
short-range spin correlations. The exponential decay of the spin-spin
correlation is produced by a finite spin gap, i. e.,  a finite energy
gap to the lowest $S=1$ excitation in the infinite ladder. Even-leg
ladders are the realization of the Resonant Valence Bond (RVB) quantum
liquid of singlets proposed by Anderson in the context of 2D $S=1/2$ AF
Heisenberg systems\cite{pwa}. 

Two-leg $S=1/2$ ladders are found in vanadyl pyrophosphate $(VO)_2
P_2O_7$ and in some cuprates like $Sr Cu_2O_3$\cite{dagrice}. Spin
susceptibility, neutron scattering, NMR and muon spin resonance
measurements are consistent with a spin gap. 

Even-leg ladders are especially interesting under doping because
theory has revealed that holes should pair to form bosons, in a
relative  "d-wave" state, with a superconducting ground-state at finite
doping\cite{hp95}.  Copper-oxide compounds exhibiting one dimensional
features with both isolated $CuO_2$ chains and $Cu_2O_3$ ladders (pairs
of $CuO_2$ chains linked by oxygen atoms between the coppers) have
recently been found to be superconducting\cite{uehara,maya}. In the
compound $Sr_2Ca_{12}Cu_{24}O_{41}$, a copper valency of $\simeq 2.2$,
i. e., a hole density of $0.2 $ per ladder-Cu is estimated \cite{maya}.
NMR and conductivity measurements have been conducted on the latter
compound, and the existence of the spin-gap is clearly established for
pressures smaller than $29$ kbar\cite{maya}. 

Thus, in the normal state of two-leg compounds the charge carriers are
likely to be bosons, restricted to move preferentially in one dimension
(1D). The proposal that CuO-planes in high-T$_c$ cuprates 
are Bose conductors in the normal state has been extensively discussed
in connection with Anderson's RVB model\cite{pwa}. However, it is  still
controversial. Besides, bosons in this context are strongly coupled to
fermionic spinons via a gauge field\cite{ilk}. The situation is clearer and simpler in the
doped spin ladders, where the bosonic carriers are not coupled to low
energy spin excitations. Materials in which these ladders occur are
good candidates for  true bosonic conductors. 

At low enough energy, a one dimensional bose gas
can be described in terms of density fluctuations with an energy
spectrum which is linear 
in wave vector\cite{haldane}. A one dimensional bose gas can thus
be mapped on a one dimensional Luttinger liquid of fermions, the
transport properties of which are known\cite{giam}: in the presence of
a random impurity potential, Anderson 
localization at low temperatures causes the resistivity to increase to
infinity as a power law of the inverse temperature. At larger
temperature a metallic behavior may step in, but no intermediate
maximum is expected. 

There is a characteristic energy scale above which the collective
density oscillation picture ceases to be valid, and the bose gas can be
described by the semiclassical quasi-particle picture\cite{lieb}.
Assume that the interaction potential between bosons can be modeled by
$c\delta(x_i-x_j)$, where $c$ is a constant. Then the excitation spectrum
remains that of independent quasi particles for $\hbar^2k^2/2m_c\gtrsim
cn_b$,  where $m_c$ is the boson mass and $n_b$ is the  linear boson density.
Thus for $k_BT>cn_b$ the 1D Bose particle transport phenomena can
be described within a semi-classical approach. At low density and small
interaction strength $c$, this temperature can be small. 
Below $k_BT \leq cn_b$ the collective mode description of the
bose gas should hold and the transport should then be described
accordingly\cite{giam}.

One of us  pointed out a long time ago \cite{gmc91} that in
Bose conductors  the condition for the validity of semiclassical transport
theory - that  the carriers mean free path is greater than their De
Broglie thermal wavelength - is much weaker than that for ordinary
(Fermi) conductors, for which the mean free path has only to be greater
than the mean distance between carriers. It was suggested in Ref.
\cite{gmc91}, on the basis 
of the Ioffe-Regel criterion \cite{ir}, that when the mean free path is
smaller than  De Broglie's thermal wavelength the resistivity saturates
at a temperature-dependent value, which may be estimated by replacing
the mean free path by De Broglie's thermal wavelength in Drude's
formula for the resistivity. 

The purpose of this note is to examine the consequences of the
semiclassical  picture on the resistivity of  
doped  quantum spin ladder systems, in the spin gap regime. The hole
pairs are treated as structureless bosonic particle: no account is
taken of the d-wave like symmetry of their wave function\cite{hp96}.
In the following  we examine the temperature dependent resistivity of a one
dimensional boson liquid interacting weakly with phonons and with
impurities. 
We propose that this straightforward analysis leads to a qualitative
understanding of the experimental behavior of the electrical
resistivity with temperature in doped quantum spin ladders\cite{uehara}.

\section{Resistivity}
\label{sec.res}

We assume that the free boson states are described by an energy band
$\epsilon({\bf k})$ the bandwidth of which for ${\bf k}$ perpendicular to the
ladders direction (c-direction) is small compared to $k_BT$. For motion
along the ladders  we assume that the bandwidth $W_c$ is large
compared to $k_BT$. Under these assumptions we may approximate
$\epsilon({\bf k})$ by $\epsilon({\bf k})=\hbar^2 k^2_c/2m_c$, where
$-\pi/a_c \leq k_c \leq \pi/a_c$  is the component of ${\bf k}$ along
the ladders direction, $a_c$ is the period within the ladder and $m_c$
is the effective mass.

The calculation of the resistivity follows the traditional Boltzmann
equation approach, as described by Ziman \cite{zim}. We consider
the contributions from boson-impurity and boson-phonon scattering to the
resistivity. 

\subsection{BOSON-IMPURITY SCATTERING}
\label{sec.bis}

The expression for the impurity scattering contribution to the
resistivity, $\rho_I$ is \cite{zim,ftn0},
\begin{equation}
\rho_I=\frac{\int d^3 k d^3 k'\,[({\bf k}- {\bf k'})\cdot {\bf u}]^2 
{\cal P}^{{\bf k}'}_{{\bf k}}}
{2k_BT\mid \int d^3k \, q \,{\bf v}_{\bf k} ({\bf k}\cdot {\bf u}) 
\frac{\partial n^{b}_{{\bf k}}}{\partial \epsilon({\bf k})} 
\mid^2} \; . 
\label{eq.rhoi}
\end{equation}
In  Eq.\ (\ref {eq.rhoi}) , ${\bf u}$ is the bosons drift velocity
along the ladders direction, ${\bf v}_{\bf k}=\partial \epsilon({\bf
k})/\partial {\bf k}$  and ${\cal P}^{{\bf k}'}_{{\bf k}}$ is the
transition  probability, given by 
\begin{equation}
{\cal P}^{{\bf k}'}_{{\bf k}}= \frac{2\pi}{\hbar}n_I\mid V_{b-i} \mid^2\,
\delta[\epsilon({\bf k}) - \epsilon({\bf 
k}')]n^{b}_{{\bf k}} (1+ n^{b}_{{\bf k}'}) \; , 
\end{equation}
where $n_I$ is the number of impurities by $cm^3$ and $V_{b-i}$ is the
boson-impurity interaction matrix element, assumed constant, 
\begin{equation}
n^{b}_{{\bf k}}=\frac{1}{e^{(\epsilon({\bf k})-\mu)/k_BT}-1}
\label{eq.bd}
\end{equation}
is the boson occupation number, and $\mu$ is the chemical potential.

The relation between $\mu$ and the number of bosons per cm  in the
ladders, $n_b$, is obtained in the usual way, assuming
non-interacting bosons\cite{hg}. We find 
\begin{equation}
n_b\lambda_c=g_{1/2}(z) \; ,
\end{equation} 
where $\lambda_c=\sqrt{2\pi/m_ck_BT}$ is De Broglie's thermal wavelength
for motion along the ladders, $z=e^{\mu/k_BT}$ and
$g_{1/2}=\sum^{\infty}_{j=1}z^j/j^{1/2}$. 

Carrying out the integrations in  Eq.\ (\ref {eq.rhoi})  we find that 
\begin{equation}
\rho_I= \frac{m_c}{nq^2}\cdot \frac{2m_cn_I\mid V_{b-i} \mid^2}{\pi
\hbar^3 n^2_b}n^{b}_{{\bf 0}}  \; ,
\end{equation}
where $n$ is the number of bosons per $cm^3$ and $n^{b}_{{\bf
0}}=z/(1-z)$ is the zero-momentum state occupation number \cite{ftn1}. 
Note that the $T$-dependence of $\rho_I$ come solely from the
$n^{b}_{{\bf 0}}$ term. The latter is a monotonously
decreasing function of $T$. The  behavior of the impurity contribution to
the resistivity with $T$ is unusual because it increases as $T$
decreases.  This is a consequence of the bosonic 1D character of the
problem. 

\subsection{BOSON-PHONON SCATTERING}
\label{sec.bps}

The boson-phonon scattering contribution to the
resistivity, $\rho_P$ is given by \cite{zim},
\begin{equation}
\rho_P=\frac{\int d^3 k d^3 k'd^3 Q [({\bf k}- {\bf k'})\cdot {\bf u})]^2 
{\cal P}^{{\bf k}'}_{{\bf k},{\bf Q}} }
{k_BT\mid \int d^3k q {\bf v}_{\bf k} ({\bf k}\cdot {\bf u}) 
\frac{\partial n^{b}_{{\bf k}}}{\partial \epsilon({\bf k})} 
\mid^2} \; , 
\label{eq.rhop}
\end{equation}
where  ${\cal P}^{{\bf k}'}_{{\bf k},{\bf Q}}$ is the transition 
probability. Assuming a deformation potential-like interaction between
the bosons and the phonons and a Debye phonon spectrum, with sound
velocity $s$, this quantity is  given by
\begin{eqnarray}
&{\cal P}^{{\bf k}'}_{{\bf k},{\bf Q}}= \frac{2\pi}{\hbar} \alpha Q 
\delta({\bf k} - {\bf k'}- {\bf Q})
\delta(\epsilon({\bf k}) - \epsilon({\bf k'})-\hbar s Q) & \nonumber \\
& \times n_{{\bf Q}} n^{b}_{{\bf k}} (1+ n^{b}_{{\bf k'}})& \;  , 
\end{eqnarray}
where $\alpha$ is a constant that characterises the boson-phonon
interaction and   $n_{\bf Q}$ is the phonon occupation number.

We write $\rho_P$ as
\begin{equation}
\rho_P = \frac{m_c}{nq^2}\cdot\frac{\alpha k^4_D}{(2\pi \hbar)^2 s
n_b}(\frac{T}{\Theta_D})^4 J(T) \; , 
\label{eq.rhopb}
\end{equation}
where  
$k_D$ is Debye's wavevector, $\Theta_D=\hbar s k_D/k_B$ is Debye's
temperature, 
\begin{equation}
J(T)= \int^{\theta_D/T}_0 d\xi \xi^4 n(\xi) \int^1_0 d\eta \eta
n(\chi_{-}(\eta))[1+ n(\chi_{+}(\eta))] \; ,
\label{eq.jt}
\end{equation}
$n(x)=(e^x-1)^{-1}$ and 
$\chi_{\pm}(\eta)=m_cs^2/2\eta^2\pm sQ/2+Q^2\eta^2/8m_c-\mu/k_BT$. 

At low temperatures, $\theta_D\gg T$, $\chi_{\pm}(\eta) \simeq
Q^2\eta^2/8m_c$ and it follows from  Eqs.\ (\ref {eq.rhopb}) and 
(\ref {eq.jt}) that  
\begin{equation}
\rho_P \propto T^3n^{b}_{{\bf 0}} \; .
\end{equation}

At high temperatures, $T \gg \theta_D$, and assuming that $m_c s^2
T/k_B\theta_D^2\ll1$ and $n_c\lambda_c \ll 1$ we find that    
\begin{equation}
\rho_P \propto  T^{1/2} \; .
\end{equation}

\subsection{ RESISTIVITY SATURATION}
\label{sec.rsat}

The expressions for $\rho$ used above,  Eqs.\ (\ref {eq.rhoi})  and
(\ref {eq.rhop}) , are valid as long as the mean free path for motion
along the ladders $\ell_c$ is greater than $\lambda_c$. For $\ell_c
<\lambda_c$ resistivity saturation is 
believed to occur. An estimate of the saturation value of the
resistivity $\rho_{SAT}$ is obtained by substituting $\ell_c$ by
$\lambda_c$ in Drude's formula  $\rho=m_c/nq^2\tau$.

Defining $\ell_c$ as $\ell_c = v_T \tau $, where $v_T$ is the bosons
thermal velocity, defined as $v_T=\sqrt{2\langle \epsilon
\rangle/m_c}$, where $\langle \epsilon \rangle$ is the bosons mean
thermal velocity. Using the boson distribution Eq.\ (\ref {eq.bd})   we find
that 
\begin{equation}
v_T=\sqrt{\frac{k_bT}{m_c}}\cdot \sqrt{\frac{g_{3/2}(z)}{g_{1/2}(z)}}\; ,
\end{equation}
where $g_{3/2}=\sum^{\infty}_{j=1}z^j/j^{3/2}$.

Thus 
\begin{equation}
\rho_{SAT}\sim \frac{m_c k_BT}{(2\pi)^{1/2}\hbar ne^2}
\sqrt{\frac{g_{3/2}(z)}{g_{1/2}(z)}} \; .
\label{eq.rhost}
\end{equation}

\section{discussion}
\label{sec.dis}

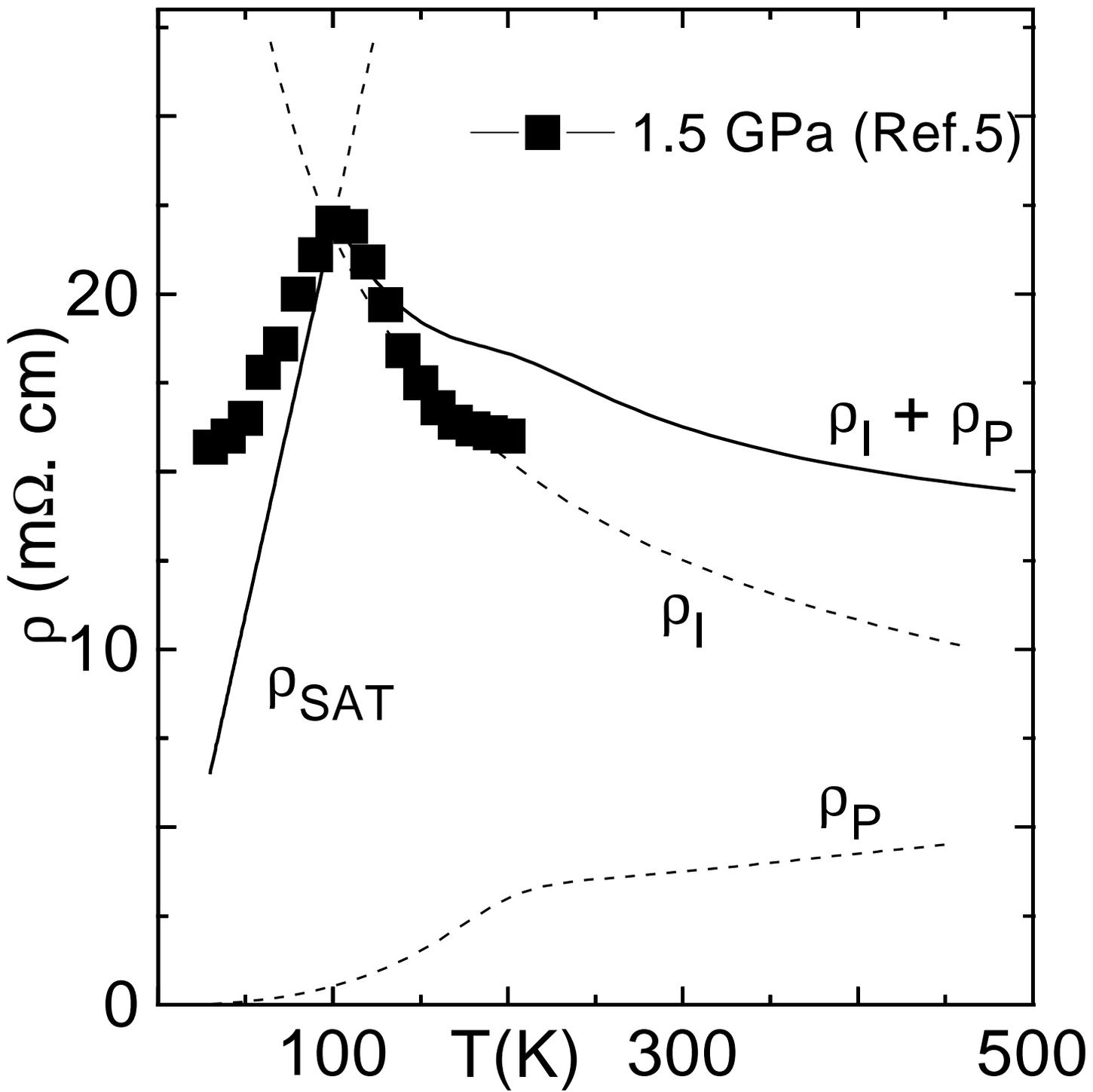
\begin{figure}
\caption{Thick line: calculated resistivity versus temperature curve
for $m_c=180m_e$ and $10^{-2}$ holes per ladder Cu. Squares:
$Sr_2Ca_{12}Cu_{24}O_{41}$ data. Dashed lines: individual contributions
from phonons, impurities and the saturation mechanism (see text).}
\label{fig.rhot}
\end{figure}

According to  Eq.\ (\ref {eq.rhost})  $\rho_{SAT}$ increases with
increasing $T$, whereas 
$\rho_I$ increases with decreasing $T$. Thus, if the boson-impurity
interaction is sufficiently strong, there is a temperature below which
$\rho_I>\rho_{SAT}$. Since, according to the Ioffe-Regel criterion the
resistivity cannot be greater than $\rho_{SAT}$,  saturation
occurs  and the resistivity curve follows $\rho_{SAT}$. Thus
there is a resistivity maximum at a temperature $T_0$ that can be
estimated as that for which $\rho_I\sim \rho_{SAT}$. This is the main
result of this paper. A typical resistivity curve predicted by our
model is shown in Fig.\ \ref {fig.rhot}. This figure also shows 
experimental resistivity data for $Sr_xCa_{1-x}Cu_{24}O_{41}$
 at $1.5$GPa obtained in Ref.\cite{uehara}.
In Fig.\ \ref {fig.rhot} the contributions to the resistivity from
phonons, impurities and from the 
saturation mechanism are  normalized  as 
$\rho_I(T=100K)=\rho_{SAT}(T=100K)=22m\Omega\cdot cm$ and
$\rho_P(T=200K) =5m\Omega\cdot cm$.  
These values for $\rho_I$ and $\rho_{SAT}$  are chosen so that
the theoretical prediction coincides with the experimental one. 
The  $\rho_P$ curve is a smooth
interpolation between the  $T\ll \theta_D$ and $T\gg
\theta_D$  behaviors, with $\theta_D=200K$. The value of $\rho_P$ at
$T=\theta_D=200K$ is chosen for illustrative purposes only.

We propose that the local resistivity maximum observed in
the  $Sr_xCa_{1-x}Cu_{24}O_{41}$ data  at $T\sim 100K$ and several
pressures \cite{uehara} arises from the same mechanism as that for the
1D boson model 
discussed above. Thus, we interpret the decreasing resistivity curves
below and above the temperature where it is maximum 
as resulting from resistivity  saturation and from boson-impurity
scattering, respectively.  In order for this interpretation to be
consistent it is necessary that, at $T\sim 100K$, $\rho_I\sim
\rho_{SAT} \sim 20m\Omega \cdot cm$. According to  Eq.\ (\ref
{eq.rhost}) , this requires that $m_c/m_en\sim 4.8\times 10^{-18}cm^3$,
where $m_e$ is the mass of the electron. 

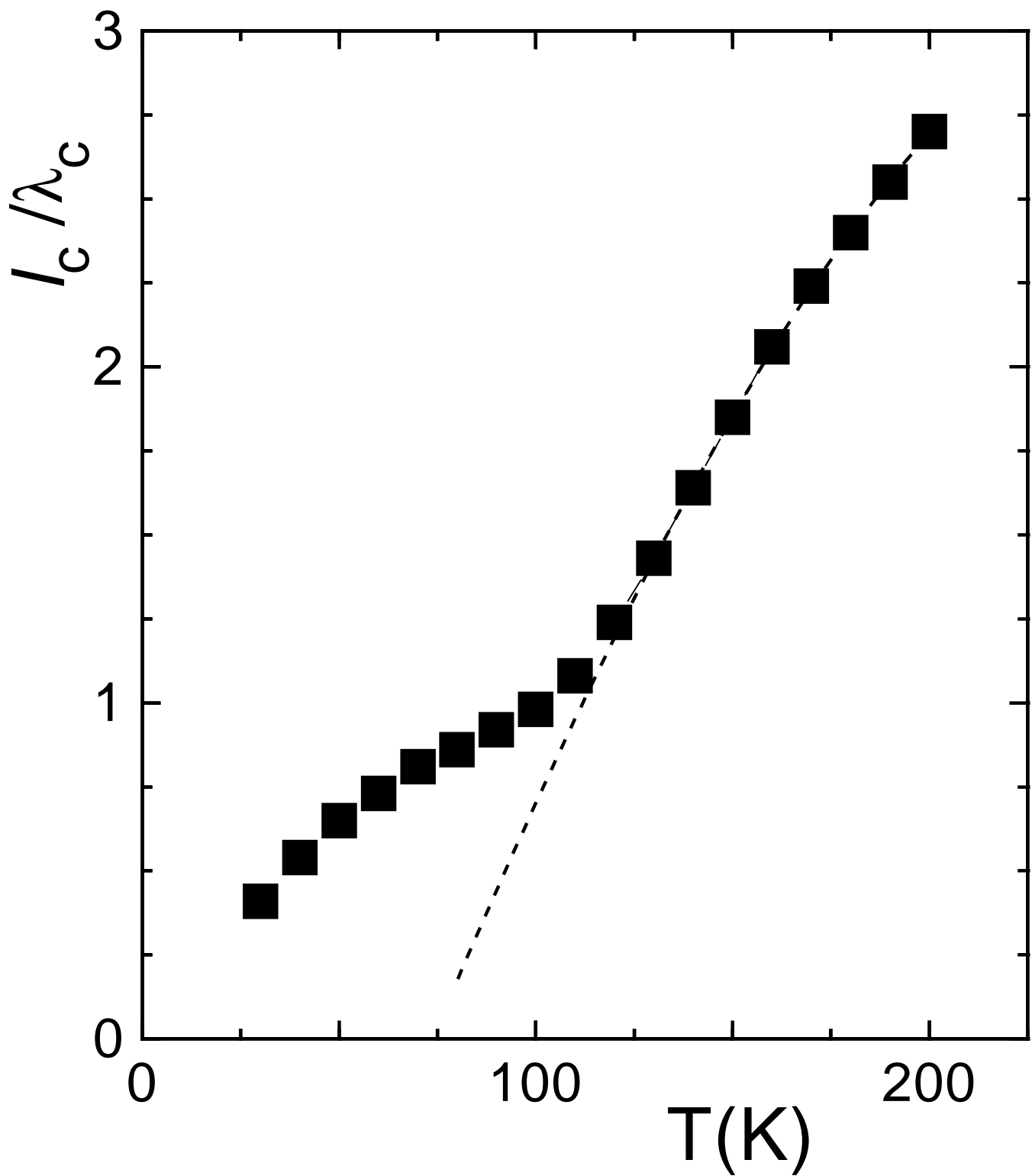
\begin{figure}
\caption{Mean free path versus temperature for $Sr_2Ca_{12}Cu_{24}O_{41}$
at $1.5$GPa . Dashed curve: extrapolation form high
temperature data.}
\label{fig.lct}
\end{figure}

We show in Fig.\ \ref {fig.lct} the mean free path along the ladder
direction obtained from  the $Sr_xCa_{1-x}Cu_{24}O_{41}$ data
at $1.5$GPa \cite{uehara} using
the Drude formula for the resistivity and the definition of $\ell_c$
given in Sec.\ \ref{sec.rsat} . In Fig.\ \ref {fig.lct} the values of
$m_c$ and $n$ are chosen so that  
$\ell_c=\lambda_c$ at $T=100K$. This requires, as discussed above,
that $m_c/m_en =4.8\times 10^{-18}cm^3$. We choose $m_c/m_e=180$ and
$n=3.7\times 10^{19} cm^{-3}$, which is equivalent to $10^{-2}$ holes
per ladder Cu atom. It is clear from Fig.\ \ref {fig.lct} that there is
a crossover around $T=100K$ from a regime of rapid variation of
$\ell_c$ with $T$, that according to our interpretation results from
boson-impurity scattering, to a regime of slow variation of $\ell_c$
with $T$, from $T\sim 100K$ to $T\sim 50K$. The latter is
consistent with the ideia of resistivity saturation, as discussed in
Ref.\cite{gmc91}.  We also note that, according to our proposal, the
boson-impurity scattering contribution to $\ell_c$ can be estimated for
$T<100K$ by extrapolating the high-temperature $\ell_c\times T$  data  to
lower $T$ values.  As shown in Fig.\ \ref {fig.lct}, this leads to  a rapidly
decreasing $\ell_c\times T$ curve. This further indicates that, within the
semiclassical picture, resistivity saturation is bound to occur in this
compound, independent of the precise value of $\ell_c/\lambda_c$ at
which it sets in. 
 
The numerical values for $m_c$ and $n$ obtained above are  rough
estimates. Neverthless, our picture requires values of $m_c$ which are 
too large compared to those obtained from theoretical estimates for
pure two-leg ladders \cite{swhite} and values of $n$ that are 
smaller than those reported in experiments ($\sim 0.2$ holes per ladder
Cu). However, the   possibility of large effective mass enhancement due
to boson-phonon interaction effects, neglected in theoretical
considerations, cannot be excluded. Besides, the Ioffe-Regel criteria
gives only  very rough order of magnitude estimates for the values of
$\ell_c/\lambda_c$ and of $\rho$ at saturation, so that the numerical
values for $m_c$ and $n$ used in our calculations may be uncertain by a
large factor.

For the semiclassical picture to be valid  a second condition must be
met: the cross over temperature 
$\theta$ to collective bosonic behavior must be smaller than the
temperature at the resistivity maximum. Our order of magnitude
estimate for this relies on the fact that the hole pair spatial
extension is $\delta$, so that the change in binding energy $\Delta E$
of a pair when colliding with another one is $\Delta E\sim
Ea_c/\delta$, because two 
hole pairs cannot occupy the same  rung. Since $E\sim 200K$,
we find   $\theta=cn_b/k_B\sim En_b\delta (a_c/\delta)\sim 20K$.
 This value of $\theta$ further supports our proposal.

To conclude, we find  that the simple minded semiclassical transport
theory proposed here for the bosonic hole pairs in the quantum spin
ladder system may account for the observed dependence of the
resistivity on the temperature observed for $Sr_xCa_{1-x}Cu_{24}O_{41}$.  
Our predictions can, in principle, be tested by changing the impurity
concentration, $n_I$. According to our semiclassical model,  increasing
$n_I$ should shift the resistivity maximum to higher temperatures and
increase its intensity, whereas decreasing $n_I$ should shift the maximum
to lower temperatures and decrease its intensity.

\acknowledgements
Work supported in part by CNPq-Bras\'{\i}lia/Brazil, Finep and FUJB.
Pascal Lederer thanks the CNPq for financial support, and C. Mauricio 
Chaves for his hospitality in PUC-Rio.

\begin{references}

\bibitem{dagrice} See, for theoretical and experimental references, the
review by E. Dagotto and T. M. Rice, Science, {\bf 271} (1996) 618. 
\bibitem{pwa}P. W. Anderson, Science {\bf 235}, (1997) 1196.
\bibitem{hp95} C. A. Hayward et al. Phys. Rev. Lett. {\bf 75}, (1995)
926. 
\bibitem{hp96}C. Hayward and D. Poilblanc, Phys. Rev. {\bf B53}, 
(1996) 11721. H.Tsunetsugu, M. Troyer, and T. M. Rice, Phys. Rev. {\bf B51},
 (1995) 16456. 
\bibitem{uehara}M. Uehara et al. J. Phys. Soc. Japan, {\bf 65}, 
(1996) 2764.
\bibitem{maya}H. Mayaffre et al. preprint (Cond. Mat. 9710063)
\bibitem{ilk} I.B. Ioffe and A. Larkin, Phys. Rev. B {\bf 39}, 
(1989) 8988.
\bibitem{gmc91}G. Carneiro, J. Phys.:Condens. Matter {\bf 3}, 
(1991) 3671.
\bibitem{ir} A. F. Ioffe and A. R. Regel, Prog. Semicond. {\bf 4}, 
(1960) 237. 
\bibitem{haldane}F. D. M. Haldane Phys. Rev. Lett. {\bf 47}, 
(1981) 1840.
\bibitem{giam}T. Giamarchi and H. J. Schultz, Phys. Rev. {B 37}, 
(1988) 325.
\bibitem{lieb}E. H. Lieb and W. Liniger Phys. Rev. {130}, (1963) 1605. 
\bibitem{zim}J.M. Ziman, {\em Electrons and Phonons}, (Oxford, London) 1967.
\bibitem{ftn0} We keep  a 3D  notation  here in order to  make contact
with the original expression derived in Ref.\cite{zim} and with 
the calculation of the boson-phonon scattering contribution, which
involves actual 3D lattice vibrations. The integral in the expression
for $\rho_I$ is, of course, effectively 1D.  
\bibitem{hg} K. Huang {\em Statistical Mechanics} (Wiley, New York) 1965.
\bibitem{ftn1} This expression is valid provided that $\mu \gg$ energy
above which the 1D Bose system excitation spectrum is quasiparticle like. 
The reason is that, according to the assumptions stated in Sec.\
\ref{sec.int} , the 
semiclassical approach is not valid below this energy so that the 
$k_c$-integrals do not extend to $k_c=0$, but have a low $k_c$ cutoff.
We check that this condition is met in the numerical calculations
performed here.
\bibitem{swhite} Steven R. White and D. J. Scalapino Phys. Rev. {\bf B 55}, 
 (1997) 6504.

\end{references}
\end{document}